\documentclass[conference]{IEEEtran}

\usepackage[T1]{fontenc}
\usepackage[]{graphicx}
\usepackage{verbatim}
\usepackage{url}
\usepackage{algorithmic}

\title{Querying Source Code with Natural Language} 

\author{\IEEEauthorblockN{Markus Kimmig}
\IEEEauthorblockA{Technische Universit\"at Darmstadt}
\and 
\IEEEauthorblockN{Martin Monperrus}
\IEEEauthorblockA{University of Lille}
\and
\IEEEauthorblockN{Mira Mezini}
\IEEEauthorblockA{Technische Universit\"at Darmstadt}
}

\begin{document}

\maketitle

\begin{abstract}
One common task of developing or maintaining software is searching the source code for information like specific method calls or write accesses to certain fields.
This kind of information is required to correctly implement new features and to solve bugs.
This paper presents an approach for querying source code with natural language. 
\end{abstract}

\section{Introduction}
\label{introduction}

The ability to correctly find source code elements is crucial for many software engineering tasks \cite{silito_murphy_volder_1}. 
For instance, a programmer may ask ``Where is the field balance read?'' before changing the way it is set, in order to avoid undesired side-effects and regression.
Providing tool-support for searching source code is especially important when the code base is huge or when  developers are unfamiliar with it.

In this paper, we present an approach that enables the developer to query source code with natural language.
Contrary to \cite{würsch_ghezzi_reif_gall_1}, our approach does not impose rules on how the queries have to be stated. It only assume that the query is grammatically correct English. It combines Part-of-Speech Tagging, statistical analysis and a touch of embedded knowledge to identify which words of the query describe what the developer is looking for.
Eventually, the natural language query is translated into a call to a code query engine.
Consequently, the scope of supported query kinds is given by the underlying code query engine, and our approach handles the multitude of variations and modulations in ways of expressing queries.
Our prototype implementation is built on top of the Eclipse JDT code query engine \cite{jdt} and the user-interface is tightly integrated with the Eclipse development environment.

The approach is evaluated with a user study of 14 subjects. 
The experiment consisted of a number of common programming and refactoring tasks which require to find an appropriate piece of code in an unknown project. 
Subjects were only allowed to use our prototype system to find the correct piece of code.
A comprehensive log of all queries from every subject shows that our system is able to correctly understand  83\% of all subject queries.
We also interviewed every subject to get detailed feedback about the prototype as well as possible future improvements. The interviews show that the tool is mostly perceived as being a valuable addition to the toolbox of software developers.

\section{From Natural Language to Formalized Queries}
\label{approach}

In our approach, when the developer has a development task that requires examining existing source code elements, (s)he expresses a query in natural language.
A translator analyzes the query and translates it to a formalized query for a code query engine.
The results given by the code query engine are displayed in the development environment, and the developer starts to examine the code elements found or refines/reformulates his query if (s)he has not found what (s)he is looking for.
It is also important to give the developer a way to inspect the result of the translation process to help him/her assess that the translation was correct.

Our approach to understanding natural language queries over source code consists of 7 sequential steps
that are presented below.
The full process is summarized figure \ref{complete-example}, along with a concrete example.

\paragraph{Cleaning and Tokenizing the Query} 

The first step is to tokenize the query. Tokenization means that the string representing the query is split into tokens which each represent a single word of the query. We use a tokenizing function based on a regular expression that splits the query string at one or more white spaces.

For example, the query
"\textit{Which methods return type integer}"
yields the following tokens:
\textit{Which}, \textit{methods}, \textit{return}, \textit{type} and \textit{integer}.
The order of the tokens is kept, because it is an important piece of information that we use in the next steps (see section \ref{parameter_values}).

\paragraph{Part-of-Speech Tagging of the Query}
\label{pos-tagging}

A Part-of-Speech (POS) Tagger is an algorithm that assigns a grammatical category (e.g. noun or verb) to every word of a sentence.
For instance, the query "\textit{Where are instances of type Integer created}" could be POS-tagged as follows: "\textit{Where(question word), are(be), instances(noun), of(preposition), type(noun), Integer(noun), created(verb)}".
POS-tagging the query enables us to enrich the query with grammatical information that we will use later for inferring the code query engine parameters.

In the following, the ordered list of important grammatical types (using only nouns and verbs) is called the grammatical form of the query.
For instance, \textit{noun -> noun -> noun -> verb} is the grammatical form of the previous query.

\begin{figure}
\includegraphics[width=\columnwidth]{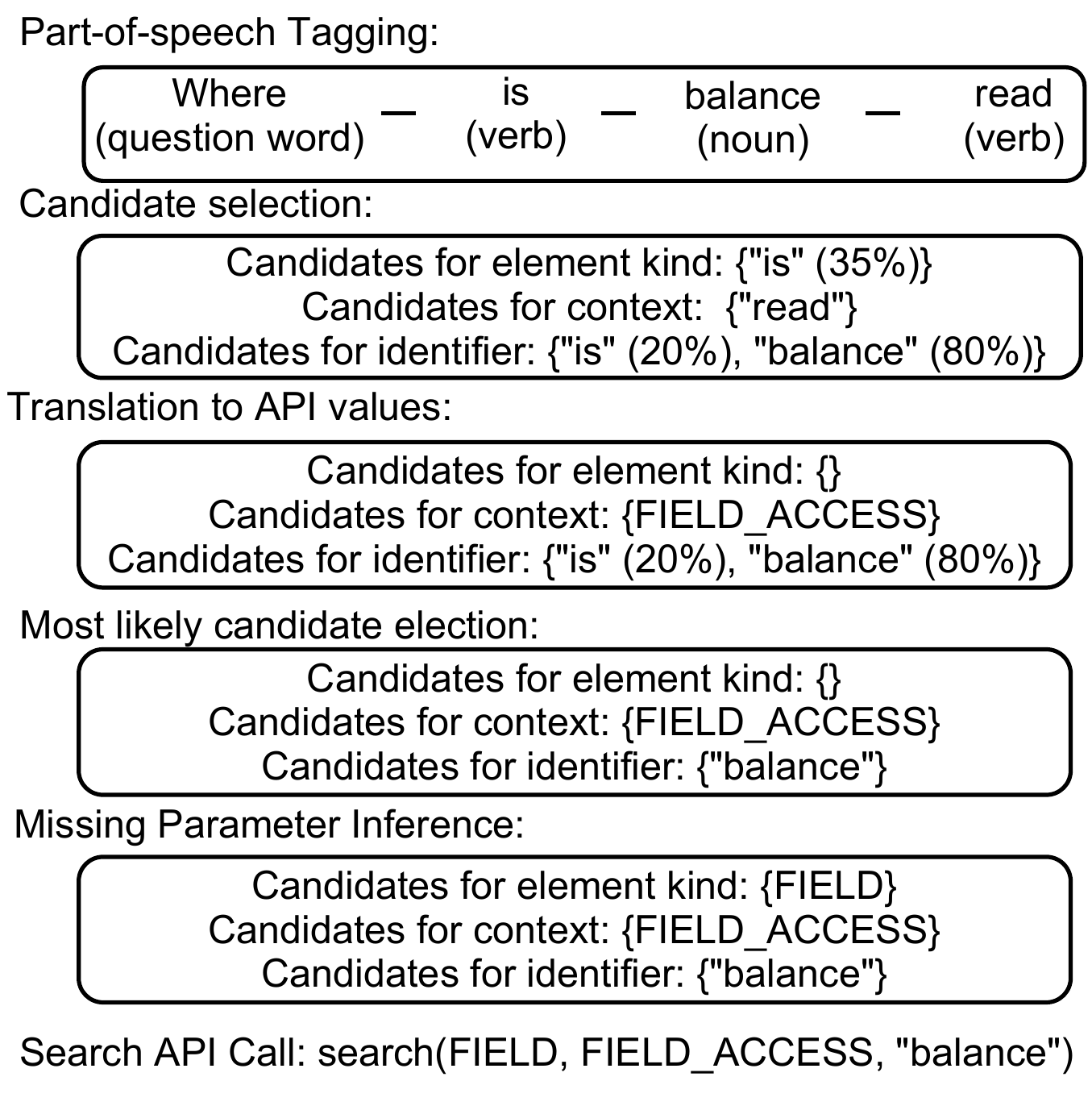}
\caption{Example Sequence of the Transformation of a Natural Language Query to an API Call with Valid Values.}
\label{complete-example}
\vspace{-.6cm}
\end{figure}

\paragraph{Selecting Code Search API Parameter Candidates}

A code query engine can have different parameters. Some parameters must take a value in a finite range of possible values, some are free.
Our prototype uses the Eclipse JDT code query engine.
The core of the JDT query engine consists of an API call taking three arguments as parameter:
the \emph{element kind} is the kind of source code element being sought (e.g. type or method);
the \emph{code context} is the programming context in which we are interested in. For instance, if the kind is ``method'', the context may be ``method call'' or ``method declaration'';
the \emph{identifier} is a arbitrary string expression describing the element that is sought (e.g. ``Integer'' or ``toStr*'').

Let us assume the user entered the query
"\textit{Which methods take a parameter of type Integer}".
We can deduce that the user is searching either a method or a type, hence they are two candidate values for the code search parameter ``element kind'': METHOD or TYPE.
For identifiers (which are free text), any word of the query would be a valid parameter value, hence the 8 words of the query are possible candidates.
We describe in this section how we use the query words, the word order and the POS-tagging information to select those candidate values.

Our selection mechanism is inspired by naive bayes classification \cite{Bishop2006}.
We first tag a set of training data to indicate which part of the query corresponds to which search API parameter, then we compute the probability of the relationship between each piece of query information and a search parameter value.

\paragraph{Training data}
\label{training-data}

The training data consists of queries which were manually annotated with information about which tokens correspond to which search API parameter.
For example, a single line from the training data is:
"\textit{What methods return:context type:kind\_of\_sourcecode\_element Integer:expression}".
It contains three annotations consisting of a pair ``word:tag'':
\emph{return:context} declares the first verb (``return'') to be the context parameter value,
\emph{type:kind\_of\_sourcecode\_element} declares the second noun (``type'') to be the kind of source code element being sought and \emph{Integer:expression} declares the third noun (``Integer'') to be the expression parameter value for this query.

Annotated queries are defined based on a combination of the implementor's expertise and real-user data collection. For replication, ours are published on \cite{replication}.
According to our experience, the number of required training data to obtain a good performance is around 250 queries. 
The reason for this relatively small number is that while the query space is infinite, the space of different grammatical query forms (as given by the POS tagger using only nouns and verbs) is  finite, and most queries can be described by a few dozens of grammatical forms.

\paragraph{Candidate Selection}
\label{candidate-selection}
When a developer enters a code search query, we first compute its grammatical form.
Then, we search the training data for annotated queries whose grammatical form (consisting only of nouns and verbs. see \ref{pos-tagging}) corresponds to the query form.
In other terms,
we search the training data for queries that match the grammatical form of the POS-tagged query.
If one is found, we add to the candidate list for the respective parameter the words from the query matching the POS-tags and positions of the accordingly annotated words from the training data.
For instance, let us assume that a training query is "\textit{Where is field::kind\_of\_sourcecode\_element  balance read?}". It contains one tuple of annotation saying that the first noun (field) corresponds to the element kind.
Let's now consider the query "\textit{Where is class Widget used?}".
Since it has the same grammatical form, we add the noun ``class'' to the candidate list for the element kind.

To decide which candidate to use, we compute a probability value for each word of the query which describes how likely this word is a candidate for the search API parameter. Note that although we use only nouns and verbs to match the query against the training data, any word of the query can be a candidate for a parameter, regardless of its POS-tag, for example the second adjective.

For instance, in the query "\textit{Which methods take a parameter of type Integer}", the grammatical form is \textit{noun -> verb -> noun -> noun -> noun}.
Let's assume that the training data contains 10 queries with the same grammatical form.
In 8 of them, the first noun refers to the code element kind.
Hence, the probability of ``methods'' to indicate the code element kind is 80\% (8/10).
However, in two queries, the third noun represents the code element kind.
In this case, the resulting candidate list for the sought element kind is: \{ method (80\%), type (20\%)\}.
To sum up, for a given query we compute NxM probabilities where N is the number of words of the query and M is the number of parameters required by the underlying code search API (e.g. M=3 for the JDT code query engine).
Probabilities are often equal to zero because many combinations have no corresponding annotations in the training queries with a matching grammatical form.

\paragraph{Translating to Concrete API Candidates}
\label{mapping}
We now have a set of candidate words for each search API parameter.
Some API parameters are free text, which means that we can pass the user query word as is.
However, most search API parameters must fit in a fixed enumeration.
For instance, the kind of sought source code elements could be either \{METHOD, CLASS, PACKAGE, ...\}.
However, the user may use different syntactical forms (e.g. plural) and synonyms. (e.g. ``function'' for ``method'').

For each search API parameter which is not free text (e.g. the element kind that is sought), we define a mapping from a list of words in stemmed form to API parameter values.
For example, if the user enters \textit{methods} (stemmed into ``method'') in the query and this word is a candidate for the source code element kind, the mapping translates it to the valid parameter value \textit{METHOD}.
Note that multiple words may be mapped to the same parameter value in order to allow the user to refer to the same programming concept with multiple words (for example \textit{class} and \textit{type}).
The mapping data is defined once at implementation time, with the keys being stemmed words and the values being the corresponding search parameter values.

Some valid API values must sometimes be deduced by a combination of words.
For example if the query contains the substring "\textit{super reference}", this hints that we are looking for method calls to the parent classes (i.e. calls to super).
We add a tuple of words to candidate lists when annotated queries contains several annotations for the same parameter.
Those tuples of words may have a corresponding entry in the mapping data that we call a multi-key.
As for simple keys, multi-keys are mapped to a single search API parameter value. 
In the previous example, there is a mapping from the tuple $<super, reference>$ to the search API parameter ``SUPER\_REFERENCE\_METHOD\_CALL''.

\begin{figure*}
\includegraphics[width=\textwidth]{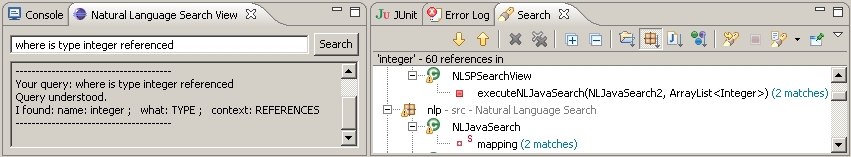}
\caption{Screenshot of our Prototype Implementation in the Eclipse IDE. The graphical interface contains a query field, a feedback area, and a search result area.}
\label{nljs_tool_screen}
\vspace{-.5cm}
\end{figure*}

\paragraph{Electing the Most Likely Search API Parameters} 
\label{parameter_values}

To elect the most likely value for each for each search API parameter, we define several rules.

\textbf{Rule \#1}:
If a candidate word has no translation to a valid parameter value in the mapping, it is removed.
For example, if \textit{Integer} is a candidate for the kind of source code element parameter, but the mapping does not contain a translation to any kind of source code element parameter value, the candidate word \textit{Integer} is removed from the candidate list for sought element kinds.

\textbf{Rule \#2}:
Sometimes candidate words are part of single key and multi-key translations. Take for example the query
"\textit{Where are parameter bounds of type Integer}".
There are two candidates for the context parameter value: \textit{parameter} indicating the API parameter value METHOD\_PARAMETER and the multi-key candidate \textit{parameter bounds} indicating the API parameter value PARAMETER\_BOUND (for example Array<I extends Integer>), which overlap in the query.
In these cases, we always choose the multi-key candidate with the highest number of words, because it is the most specific.

\textbf{Rule \#3}
If several candidates remain after rule \#1 and \#2, the chosen API parameter value is the one with the highest probability computed from the training queries (see \ref{candidate-selection}).
If no value could be found, the value is set to unknown.

\textbf{Rule \#4}:
We always start by electing the context parameter value, then we elect the code element kind, and then the identifier.
Starting with the context parameter has two rationales. First, according to our experience, this parameter is correctly chosen with the highest reliability. Second, it enables us to infer the correct element kind is many cases, as shown below in \ref{missing-values-inference}.

\textbf{Rule \#5}:
If a candidate has been chosen as the final parameter value, all other candidates with the same token are removed from the other candidate lists.
For example, let us assume that the element kind candidate list consists of the first noun, and that the context candidate list contains the first noun and the second verb.
If the first noun is elected as the value for the context parameter, the first noun is removed from all other candidate lists.

\textbf{Rule \#6}:
The last parameter value to choose is the identifier (a free string). 
We first check whether a word is indicated as such with double quotes in the input query (e.g. \emph{Where is declared metho ''printToConsole``?}). If this is the case, the identifier parameter is set to this quoted word. If this is not the case but there is exactly one word in the query which contains wildcard characters (* or ?), this word is chosen. Otherwise the word with the highest probability to be correct is chosen (see rule \#3).

\paragraph{Inferring Missing Values}
\label{missing-values-inference}
At this point, for each search API parameter, we have either a unique valid value or nothing.
To be able to process queries that provide incomplete information, we also store the information of the dependency between search API parameter values.
For instance, the context parameter often implies one specific kind of source code element to be sought.
Take for example the following query:
``\textit{Where is number read}''.
This query does not contain any explicit information on the kind of source code element that is sought (e.g. type, method, etc.). However, the word ``read'' indicates the context parameter value \textit{READ\_ACCESS}. With this information, we are able draw the conclusion that the only possible kind of source code element parameter value is \textit{FIELD}, because all other code elements, like for example \textit{package} or \textit{method}, can not be read.

\paragraph{Performing the Code Search}

Figure \ref{complete-example} sums up our algorithm to translate a natural language query into a set of valid code query engine parameter values.
The final step eventually consists of using the target search API to carry out the search request with the parameter values determined by the presented strategy. The search results are then displayed to the developer.
Our prototype is integrated into the Eclipse IDE as a plug-in. A screenshot is shown in figure \ref{nljs_tool_screen}.

\section{Evaluation}
\label{evaluation}
To evaluate our system, we conducted a user experiment.
We recruited 14 subjects to perform 13 code maintenance tasks which require navigating over source code (e.g. \emph{``Method init() is called in a method where it doesn't make any sense. Remove the method call.})''. 
The experiment takes about 30 minutes per subject, and consists of three phases. First, the experimenter presents the experiment and the prototype to the subject. Then, the subject works on the given tasks for about 15 to 20 minutes. 
The experiment concludes with an oral and recorded interview. Subjects are students in computer science at our university

The subjects can only use the Eclipse IDE to complete the tasks. The only IDE feature they are allowed to use is the natural language search tool.
In particular, to browse and find code, they are not allowed to use the ``package explorer view'' and the ``outline view'', the ``open type widget'', the text and java search widgets. Indeed, the goal of the experiment is to find out how well our approach is sufficient for finding code elements.

From all 14 sessions, we computed the number of understood queries and the number of correctly understood queries. These have to be separated, because the fact that the tool understood a query only means it found a value for all three parameters, but not necessarily the right ones.
For example, if the user enters the query "\textit{Where is method init() called}" the tool might choose \textit{method} for the ``search for'' parameter, \textit{call} for the context parameter and \textit{where} for the expression parameter, which is wrong. The correct values would be be {METHOD, CALL, "init()"}.
The average percentage of understood queries was 91\% (251/276). The average percentage of correctly understood queries was 83\% (228/276). 
However, about 91\% (229/251) of all understood queries were understood correctly.
We had 5 subjects had previous experience with the Eclipse code search widget, 4 of them stated they would prefer working with our code search system rather than the Eclipse one.
As future work, we plan a controlled experiment to thoroughly compare both systems.

\section{Related Work}
\label{related-work}

A large body of work has been done on developing systems and frameworks to query source code.
There are several approaches that use specialized query languages to retrieve information about source code (e.g., \cite{hajiyev_verbaere_moor_1,janzen_volder_1}. These approaches require learning the syntax of the query language before being able to work with the tools, while with a natural language interface there is almost no learning phase.

Queries are not only over source code.
For instance, de Alwis and Murphy \cite{alwis_murphy_1} described different software maintenance queries.
Hill et al. \cite{Hill2009} uses NLP based on program identifiers to improve contextual code search  (which pieces of code are about this topic?).
Ko et al. \cite{Ko2008} presents a system for querying program output, and not source code itself.
Wang et al.'s approach \cite{Wang2008} detects duplicate bug reports using NLP.

W\"ursch et al. \cite{würsch_ghezzi_reif_gall_1} described a powerful system that is similar to ours.
Our technical contribution describes a completely different algorithm using incomparable techniques.
While they use tools from ontology engineering, we use natural-language processing techniques.
Our user study also contributes with first insights on how developers react on using such systems.
But apart from these technical differences, our approach is novel in the sense that it supports completely free queries, while theirs is based on guided input, i.e. the developers select a query into an adaptive list of possible queries. This new degree of freedom creates a whole new challenge, and our paper aims at contributing to this new research direction.

\section{Conclusion}
\label{conclusion}
We presented our approach for querying source code with natural language queries.
The approach translates natural language queries to concrete parameters of a third party code query engine. Our approach uses a combination of natural language processing techniques (Part-Of-Speech tagging, stemming), as well as a custom algorithm that extracts statistical data from manually annotated training queries. Our prototype implementation uses as underlying code query engine an unmodified off-the-shelf version of the Eclipse JDT code query engine.
We evaluated our approach using a user-study with a total of 14 subjects.
Our prototype was able to correctly understand 83\% of queries: 91\% of 276 queries which have been entered by subjects were recognized and 91\% of them correctly. Future work consists of conducting a controlled experiment to compare our approach against related systems on the same set of tasks.

\bibliographystyle{ieeetr}
\normalsize
\bibliography{bibliography}

\end{document}